\documentclass{MITcsail}
\usepackage[english]{babel}
\usepackage[letterpaper,top=2cm,bottom=2cm,left=3cm,right=3cm,marginparwidth=1.75cm]{geometry}
\usepackage{breakurl}        % For breaking URLs easily trough lines
\providecommand{\keywordsname}{Keywords}
\long\def\keywords#1{%
    \gdef\@keywords{\message{\keywordsname}%
        {\abstractsize\noindent{\keynamefont
      \keywordsname:~}#1\par \vskip.7\baselineskip}}}

\chardef\us=`\_

\title{Association between a Failed Prominence Eruption and the Drainage of Mass from Another Prominence}
\runningtitle{Association between Two Prominence Eruptions}

\author
{Jianchao Xue~$^{1}$\footnote{E-mail: xuejc@pmo.ac.cn}, Li Feng~$^{1,2}$, Hui Li\footnote{Correspondence E-mail: nj.lihui@pmo.ac.cn}~$^{1,2}$, Ping Zhang~$^{1}$, Jun Chen~$^{1}$, Guanglu Shi~$^{1,2}$, Kaifan Ji~$^{3}$, Ye Qiu~$^{4}$, Chuan Li~$^{4,5}$, Lei Lu~$^{1}$, Beili Ying~$^{1}$, Ying Li~$^{1,2}$, Yu Huang~$^{1}$, Youping Li~$^{1}$, Jingwei Li~$^{1}$, Jie Zhao~$^{1}$, Dechao Song~$^{1}$, Shuting Li~$^{1,2}$, Zhengyuan Tian~$^{1,2}$,  Yingna Su~$^{1,2}$, Qingmin Zhang~$^{1}$, Yunyi Ge~$^{1}$, Jiahui Shan~$^{1,2}$, Qiao Li~$^{1,2}$, Gen Li~$^{1,2}$, Yue Zhou~$^{1,2}$, Jun Tian~$^{1,2}$, Xiaofeng Liu~$^{1,2}$, Zhichen Jing~$^{1,2}$, Bo Chen~$^{6}$, Kefei Song~$^{6}$,   Lingping He~$^{6}$, Shijun Lei~$^{1}$, Weiqun Gan~$^{1,7}$\\
\vspace{1em} % Space between authors and afilliations
\normalfont{\small $^{1}$Key Laboratory of Dark Matter and Space Astronomy, Purple Mountain Observatory, Chinese Academy of Sciences, 10 Yuanhua Road, Nanjing 210023, People's Republic of China}\\
\normalfont{\small $^{2}$School of Astronomy and Space Science, University of Science and Technology of China, 96 Jinzhai Road, Hefei 230026, People's Republic of China}\\
\normalfont{\small $^{3}$Yunnan Observatories, Chinese Academy of Sciences, Kunming 650216, People's Republic of China}\\
\normalfont{\small $^{4}$School of Astronomy and Space Science, Nanjing University, Nanjing 210023, People's Republic of China}\\
\normalfont{\small $^{5}$Key Laboratory for Modern Astronomy and Astrophysics (Nanjing University), Ministry of Education, Nanjing 210023, People's Republic of China}\\
\normalfont{\small $^{6}$Changchun Institute of Optics, Fine Mechanics and Physics, Chinese Academy of Sciences, 3888 Dong Nanhu Road, Changchun 130033, People's Republic of China}\\
\normalfont{\small $^{7}$University of Chinese Academy of Sciences, Nanjing 211135, People's Republic of China}\\
}

\begin{document}

\maketitle

\begin{abstract}
    \textbf{Abstract:} Sympathetic eruptions of solar prominences have been studied for decades, however, it is usually difficult to identify their causal links.
            Here we present two failed prominence eruptions on 26 October 2022 and explore their connections.
            Using stereoscopic observations, the south prominence (PRO-S) erupts with untwisting motions, flare ribbons occur underneath, and new connections are formed during the eruption. The north prominence (PRO-N) rises up along with PRO-S, and its upper part disappears due to catastrophic mass draining along an elongated structure after PRO-S failed eruption.
            We suggest that the eruption of PRO-S initiates due to a kink instability, further rises up, and fails to erupt due to reconnection with surrounding fields. The elongated structure connecting PRO-N overlies PRO-S, which causes the rising up of PRO-N along with PRO-S and mass drainage after PRO-S eruption. This study suggests that a prominence may end its life through mass drainage forced by an eruption underneath.
            % Evidences supporting breakout reconnection are explored.
\end{abstract}

\section{Introduction}\label{s:intro}
    % Filaments and their physical paramenters
    % Filament eruptions and recent explanations, oscillation and mass drainage, ideal instabilities, reconnection
    % Sympathetic eruptions
    % Introduce recent achievements in explaining filament eruption.
    %______________________________________________________________
    Solar prominences, or filaments when they are seen on the disk, are dense
    structures extending into the solar corona \citep{2015ASSL..415.....V}. Cold and
    dense prominence mass is generally thought to be suspended by the magnetic tension
    force in magnetic dips \citep{1957ZA.....43...36K,1974A&A....31..189K}. Based on the
    statistics of \citet{2017ApJ...835...94O}, 11\% of filaments have a magnetic
    flux rope configuration, while others appear as sheared arcades. A magnetic flux rope
    refers to a set of magnetic field lines winding around a common axis more
    than once \citep{2017ScChD..60.1383C}. They may suffer from ideal
    magnetohydrodynamic (MHD) instabilities, mainly a kink instability
    \citep{1981GApFD..17..297H} and a torus instability
    \citep{2006PhRvL..96y5002K}. These instabilities cause eruptions accompanied by large solar flares and coronal mass ejections (CMEs).
    For a uniformly twisted flux rope, the kink instability sets if the twist number exceeds 1.25 \citep{1981GApFD..17..297H}. The torus instability sets if the external poloidal field decreases fast enough, which is evaluated by a decay index. In theory, the critical value of this index for a circular and a straight flux rope is 1.5 and 1, respectively \citep{2010ApJ...718.1388D}. 
    % The equilibrium of a prominence is mainly achieved through downward gravitational force, net upward Lorentz force, and less important gas pressure gradient \citep{2023ApJ...956..106W}. Among them, the Lorentz force consists of hoop force, strapping force, tension force, and non-axisymmetry (of a flux rope) induced forces \citep{2021NatCo..12.2734Z}. When the equilibrium is destroyed due to ideal magnetohydrodynamic (MHD) instabilities, mainly helical kink instability \citep{1981GApFD..17..297H} and torus instability \citep{2006PhRvL..96y5002K}, or magnetic reconnection, the prominence may erupt accompanying or without solar flare and/or coronal mass ejection (CME). 

    Solar filament eruptions are generally classified into full eruptions, partial
    eruptions, and failed or confined eruptions
    \citep{2007SoPh..245..287G}. The former two types develop
    into CMEs with significant amounts of high-energy particles and magnetic plasma
    ejected into the interplanetary space. But for a failed eruption \citep{2003ApJ...595L.135J}, neither the prominence mass nor the supporting magnetic structure escapes after an initial
    acceleration. Causes for the failed eruption of a flux rope can be
    classified into four types \citep{2023ApJ...956..106W,2023ApJ...959...67C}:
    (1) not reaching the torus instability, where the eruption initiates due to a kink instability \citep[failed kink regime, ][]{2005ApJ...630L..97T}, or the decay index has a saddle-like profile \citep{2010ApJ...725L..38G}; (2) failed-torus regime, where a torus-unstable
    flux rope fails to erupt due to a downward tension force induced by the
    external toroidal field \citep{2015Natur.528..526M,2023ApJ...956..106W} or a
    downward Lorentz force by a non-axisymmetry of the flux rope
    \citep{2021NatCo..12.2734Z,2023arXiv231207406Z}; (3) magnetic reconnection-caused destruction of the flux
    rope in a breakout-like configuration, where a quadrupolar field is necessary
    \citep{2008ApJ...680..740D,2012A&A...548A..89N,2023ApJ...951L..35C,2023ApJ...959...67C};
    (4) magnetic reconnection-caused destruction due to the flux rope writhing, the process may happen between the flux rope and the external field, or
    between the flux rope legs, where the magnetic field configuration could be dipolar
    \citep{2016ApJ...832..106H,2019ApJ...878...38Y,2019ApJ...877L..28Z,2023MNRAS.525.5857J}. In the
    regimes mentioned above, an eruption initiates in general, though not always, due to MHD
    instabilities and further accelerates with flare reconnection below the flux rope
    \citep{2012ApJ...760...81K,2023ApJ...956..106W}. In the breakout configuration,
    the breakout reconnection initially promotes the eruption through removing the
    strapping field. However, when the magnetic reconnection between the external field and the flux rope starts, the flux rope is destroyed and upward hoop force decreases, which results in the failure of the eruption \citep{2023ApJ...951L..35C}.
    % Though some authors thought that an eruption could be triggered by reconnection and the impulsive phase is driven by torus instability \citep{2013ApJ...769L..25C}.
    % In the breakout configuration, which is mainly concerned in this work, the breakout reconnection experiences two stages \citep{2023ApJ...959...67C}. One is that the flux above the flux rope from the inner polarities involves reconnection, which promotes the flux rope eruption in combination with the reconnection below the flux rope, as the breakout model proposed by \citet{1999ApJ...510..485A}. In the second stage, the flux rope itself involves reconnection, which erodes the poloidal flux and decreases the upward hoop force. Failed eruption observations supporting the breakout configuration can be found in the works of \citet{2012A&A...548A..89N} and \citet{2023ApJ...959...67C}. 
    % the flux rope is ruined by the reconnection with external field \citep{2008ApJ...680..740D,2023ApJ...951L..35C,2023MNRAS.525.5857J}. In the breakout configuration, a quadrupolar field is necessary, and 

    Some eruptions may trigger another eruption, or have a common origin.
    Solar sympathetic (homologous) eruptions refer to the eruptions
    that occur at different (the same) sites during a relatively short interval
    with a certain physical connection \citep{2003ApJ...588.1176M}. However for
    sympathetic eruptions, it is generally difficult to determine whether there is just a chance coincidence.
    Since solar eruptions in the corona are often dominated by the magnetic field, the causal links between the sympathetic eruptions
    are generally thought to be of a magnetic nature \citep{2021ApJ...912L..15T}.
    Magnetic reconnection at separatrices, separators, and
    quasi-separatrix layers is the most common cause for sympathetic eruptions in
    several works
    \citep{2009ApJ...703..757L,2011ApJ...739L..63T,2018ApJ...869..177W,2021ApJ...912L..15T}.
    Magnetic reconnection, along with the expansion of an erupting flux rope
    \citep{2012ApJ...750...12S,2023ApJ...943...62Y}, mainly plays a role in eroding
    the overlying field that provides a confining force
    \citep{2009ApJ...703..757L,2011ApJ...739L..63T,2013ApJ...769L..25C,2017ApJ...844...70L}, and
    exchanging magnetic flux between flux ropes \citep{2018ApJ...869..177W}.
    Other authors suggest that perturbations caused by surges, CMEs, and waves can also
    result in sympathetic eruptions
    \citep{2001ApJ...559.1171W,2011ApJ...738..179J,2021ApJ...923...74D}.
    % Generally sympathetic eruptions consists of successful eruptions, however to the best of our knowledge, association between two failed eruptions have not been reported.  
    % expansion during filament eruption or by evacuation after the eruption can also result in sympathetic eruptions \citep{2012ApJ...750...12S}
    % Sympathetic eruptions are linked by disturbance caused by a surge \citep{2001ApJ...559.1171W}.

    In this work, we report two prominence eruptions on 26 October 2022, where a
    failed eruption is followed by the catastrophic drainage of mass of another
    prominence. We analyze the causes of the two eruptions and explore their
    associations. We introduce the observations in
    Section~\ref{s:obs}. Since this is the first light from the Lyman-alpha Solar
    Telescope \citep[LST, ][]{2019RAA....19..158L,2019RAA....19..159C,2019RAA....19..162F} onboard the
    Advanced Space-based Solar Observatory \citep[ASO-S, ][]{2023SoPh..298...68G},
    emissions in H~\textsc{i} Ly$\alpha$, He~\textsc{ii}
    $30.4\,\mathrm{nm}$, and H~\textsc{i} H$\alpha$ lines are compared briefly. In
    Section~\ref{s:res}, the erupting processes of the two
    prominences are shown. Their eruption causes and association are discussed in
    Section~\ref{s:dis}.

    \section{Observations and Analysis}\label{s:obs}
    \begin{figure}
        \centerline{\includegraphics[width=1\textwidth,clip=]{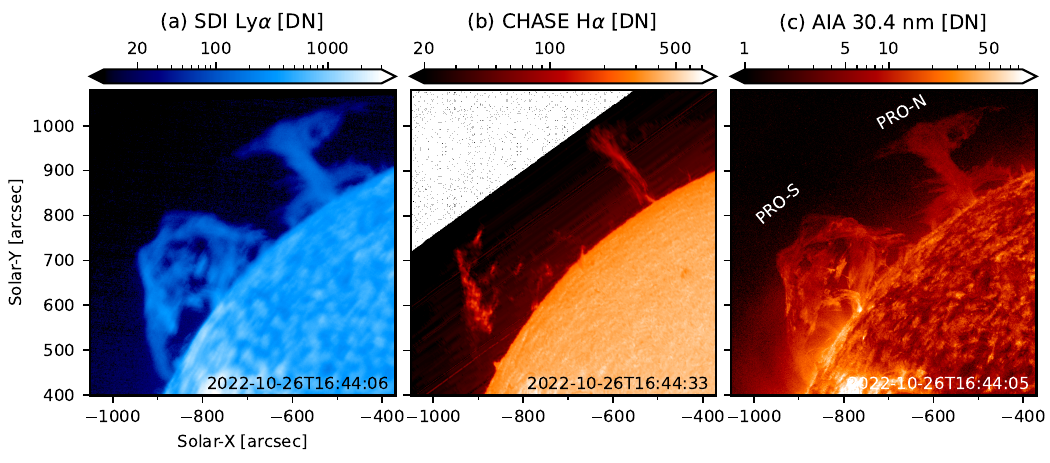}}
        \caption{\label{fig-ove}Prominence images of SDI Ly$\alpha$, CHASE H$\alpha$, and AIA $30.4\,\mathrm{nm}$. An animation of SDI, CHASE H$\alpha$ line center, and H$\alpha$ Doppler images from 16:41 to 19:00~UT is attached as supplementary material, see movie\_a.mp4. For the animation of AIA 30.4, 17.1, and $19.3\,\mathrm{nm}$ observations from 16:26 to 20:30~UT, see movie\_b.mp4. } %
    \end{figure}

    On 26 October 2022, two solar prominences on the northeastern limb erupted sequentially. Neither flare (by checking GOES soft X-ray light curve) nor associated CME (in the field-of-view of the SOHO/LASCO coronagraph) were found. They were observed by the ASO-S/LST during the commissioning phase. The LST consists of three instruments: a solar disk imager (SDI) in {H~\textsc{i}} Ly$\alpha$ (Ly$\alpha$ for short), a white-light solar telescope (WST) in $360\pm 2\,\mathrm{nm}$ waveband, and a solar corona imager (SCI) in both Ly$\alpha$ and white light ($700\pm 32\,\mathrm{nm}$). SCI started working in December 2022, hence SCI observations are not available. SDI has a pixel scale of $0.5\,\mathrm{arcsec}$, however, the spatial resolution is worse than that of the Atmospheric Imaging Assembly \citep[AIA, ][]{2012SoPh..275...17L}, i.e., $1.5\,\mathrm{arcsec}$. The SDI cadence is $10\,\mathrm{seconds}$ during first light, and data between 15:44~UT and 16:41~UT are lost. SDI level-1 data are dark-field and flat-field corrected. They are aligned to AIA $30.4\,\mathrm{nm}$ using cross-correlation and optical flow methods \citep{2022RAA....22f5010C}.
    %They are the first sympathetic eruption event, if they are sympathetic eruptions but not a temporal coincidence, observed by the LST onboard the ASO-S mission during the commissioning phase.
    % After their eruption, no associated CME is found in the field-of-view (FoV) of the Large Angle and Spectrometric Coronagraph \citep[LASCO, ][]{1995SoPh..162..357B} onboard the SOHO mission.
    %Lyman-alpha Solar Telescope \citep[LST, ][]{2019RAA....19..158L,2019RAA....19..162F} onboard the Advanced Space-based Solar Observatory \citep[ASO-S, ][]{2023SoPh..298...68G}

    This event was also observed by the Chinese H$\alpha$ Solar Explorer \citep[CHASE, ][]{2022SCPMA..6589602L,2022SCPMA..6589603Q} H$\alpha$ Imaging Spectrograph (HIS). In raster scanning
    mode, HIS scans the solar disk in $1.2\,\mathrm{min}$ with a pixel size of
    $1.04\,\mathrm{arcsec}$ and spectral resoltion of $0.024\,\mathrm{\AA}$
    ($0.048\,\mathrm{\AA}$ in binning mode for this observation). CHASE data are
    dark-field, slit-image-curvature, and flat-field corrected; wavelength
    calibration is conducted using two photospheric absorption lines in a quiet
    region around solar disk center. CHASE maps are co-aligned with the Helioseismic and Magnetic Imager \citep[HMI, ][]{2012SoPh..275..207S} continuum
    images using the scale-invariant feature transform \citep[SIFT,][]{lowe2004,
        2017Ji}. Figure~\ref{fig-ove} shows snapshots of the prominence eruptions in
    SDI Ly$\alpha$, CHASE H$\alpha$, and AIA $30.4\,\mathrm{nm}$. Though SDI has a
    lower spatial resolution than AIA $30.4\,\mathrm{nm}$ images, SDI has larger
    counts, hence a higher signal to noise ratio.

    The Ly$\alpha$ lines of both H~\textsc{i} at $121.6\,\mathrm{nm}$ and
    He~\textsc{ii} at $30.4\,\mathrm{nm}$ are among the brightest lines of the
    solar spectrum. On the basis of irradiance and imaging observations, it is
    found that their intensities have a close relationship
    \citep{2005ApJ...622..737A,2022A&A...657A..86G}. For the emission of
    $121.6\,\mathrm{nm}$ and $30.4\,\mathrm{nm}$ lines from prominences, resonant
    scattering of chromspheric radiation is an important contributor
    \citep{2022A&A...665A..39Z}. Compared with them, the H~\textsc{i} H$\alpha$ line at
    656.28\,nm is relatively optically thin and mainly contributed from the prominence
    core \citep{1993A&AS...99..513G}. This is the main reason why the prominences in Figure~\ref{fig-ove} look
    thinner in H$\alpha$ than those seen in Ly$\alpha$ and $30.4\,\mathrm{nm}$
    images. Using a non-LTE radiative transfer simulation,
    \citet{1993A&AS...99..513G} found that the H$\alpha$ brightness is mainly
    determined by the prominence emission measure ($\mathrm{EM}=\int n_e^2 \mathrm{d}z$
    where $n_e$ is the electron density and $z$ is distance along the line-of-sight). Hence
    the left (south) leg of the south prominence, PRO-S in abbreviation, should
    have a larger EM than its spine and right leg during its eruption. This
    brightness distribution suggests that the PRO-S eruption is asymmetric. %, which is possibly associated with that the right side erupts faster than the left side, left leg roots in an active region (AR, see Figure~\ref{fig-eui_phi}), left leg is tight seen in Figure~\ref{fig-eui_proS} is tight than the right leg.

    \begin{figure}
        \centerline{\includegraphics[width=1\textwidth,clip=]{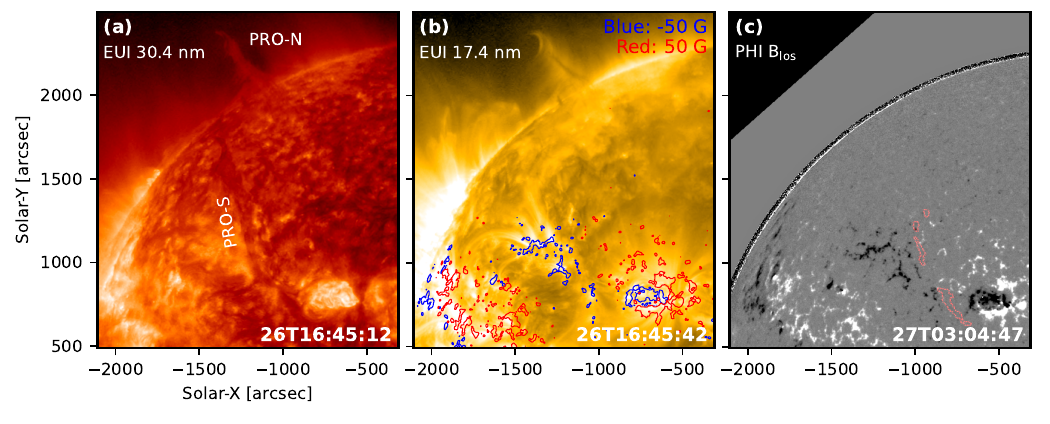}}
        \caption{\label{fig-eui_phi}Maps in Solar Orbiter view. (a)--(b) EUI-FSI 30.4 and $17.4\,\mathrm{nm}$ maps at 16:45~UT on 26 October 2022. (c) PHI $B_\mathrm{los}$ at 3:04~UT on 27 October 2022. The blue and red contours in (b) represent $B_\mathrm{los}$ of -50 and $50\,\mathrm{G}$, respectively. The contours in (c) mark the PRO-S location, which are extracted from EUI $30.4\,\mathrm{nm}$ at 2:00~UT on 26 October 2022. An animation of EUI 30.4 and $17.4\,\mathrm{nm}$ observations from 10:05~UT on 26 October to 4:25~UT on 27 October is available as supplementary material, see movie\_c.mp4.}
    \end{figure}
    With respect to Earth view, the Solar Orbiter \citep{2021A&A...646A.121G}
    is at the eastern side with an angle of around $46^\circ$. The prominences
    observed by the Full Sun Imager (FSI) of the Extreme Ultraviolet Imager
    \citep[EUI, ][]{2020A&A...642A...8R,2023eui_release} are shown in Figure~\ref{fig-eui_phi},
    where the erupting PRO-S is seen as a solar filament, and the north prominence,
    PRO-N, is still above the solar limb. EUI-FSI 30.4 and $17.4\,\mathrm{nm}$ images
    have a pixel scale of $4.44\,\mathrm{arcsec}$, corresponding to
    $1.32\,\mathrm{Mm\,pixel^{-1}}$ as the distance of Solar Orbiter from the Sun
    is $\approx 0.41\,\mathrm{AU}$. EUI-FSI has a lower cadence of 10 or 20\,minutes
    than AIA EUV images of 12\,seconds. EUI $17.4\,\mathrm{nm}$ images are mainly
    contributed by the emission from Fe~\textsc{x} and Fe~\textsc{ix}, similar to AIA
    $17.1\,\mathrm{nm}$ channel dominated by Fe~\textsc{ix} lines with characteristic
    emission temperature of 0.63\,MK. There is no observation from the EUI High
    Resolution Imager (HRI). Observations of the Polarimetric and Helioseismic
    Imager \citep[PHI, ][]{2020A&A...642A..11S} related to this event started from
    27 October 2022 at 3:04~UT. A snapshot of PHI line-of-sight magnetic field
    ($B_\mathrm{los}$) is shown in Figure~\ref{fig-eui_phi}c, where white (black)
    represents positive (negative) polarities, and the overlaid contours mark the
    PRO-S location at 2:00~UT on 26 October 2022 extracted from an EUI
    $30.4\,\mathrm{nm}$ image. PHI has a pixel scale of $3.57\,\mathrm{arcsec}$, and the vector magnetogram from PHI is not available for this event.
    The difference in light travel time between Solar Orbiter and Earth-orbit
    telescopes is corrected and the time at Earth is used in this article.
    Figure~\ref{fig-eui_phi} shows that PRO-S is an intermediate filament and PRO-N
    is a polar-crown one.

    %%%%%%%%%%%%%%%%%%%%% Eruption %%%%%%%%%%%%%%%%%%%%%%%
    \section{Results}\label{s:res}
    \subsection{Time-Distance Diagrams of the Erupting Prominences} \label{subs:td}
    \begin{figure}
        \centerline{\includegraphics[width=1\textwidth,clip=]{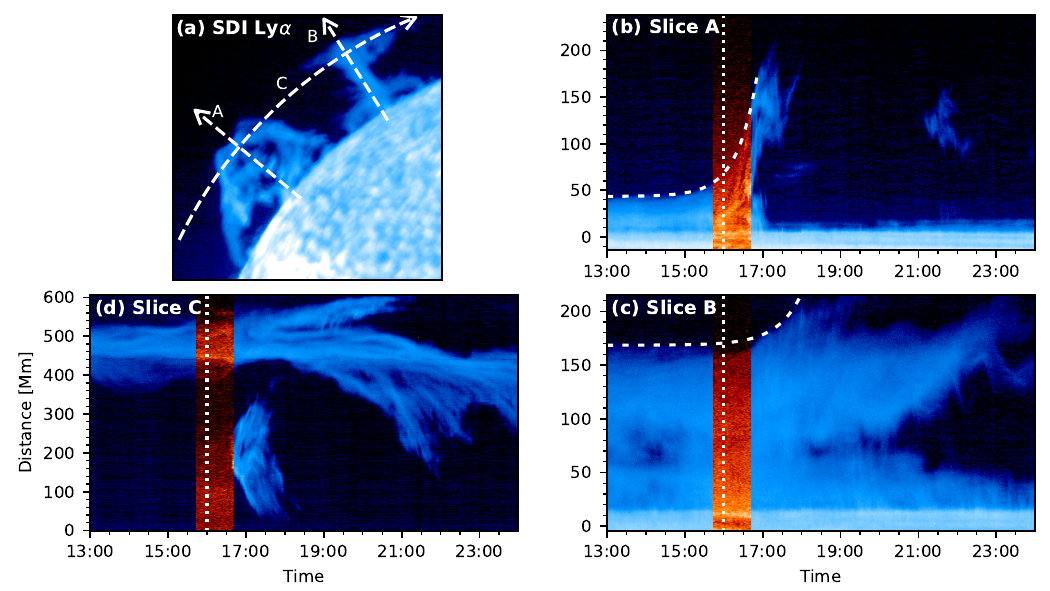}}
        \caption{Time-distance diagrams of prominence eruptions. (a) SDI map with the location of artificial slices indicated, and arrows representing directions of the slices. (b)--(d) Time-distance diagrams along slices A--C, the blue parts are from SDI images and the red parts are AIA $30.4\,\mathrm{nm}$ observations.\label{fig-slice}}
    \end{figure}

    To trace the dynamical evolution of the prominences, time-distance diagrams are
    synthesized along three artificial slices (Figure~\ref{fig-slice}). Among them,
    slices A and B are perpendicular to the solar limb and almost along the rising
    trajectories of PRO-S and PRO-N, respectively. Slice C is parallel with the
    solar limb. Considering that SDI images have a larger signal-to-noise ratio, we
    mainly used SDI observations and the missing parts are replaced by AIA
    $30.4\,\mathrm{nm}$ images (shown in blue and red respectively in
    Figures~\ref{fig-slice}b--d). The height variations of PRO-S and PRO-N, with
    respect to the solar limb, are fitted using a function composed of a linear
    part and an exponential part \citep[][dashed lines in Figure~\ref{fig-slice}b--c]{2013ApJ...769L..25C}.
    The rising speeds of PRO-S and PRO-N are calculated to be around 70 and
    $45\,\mathrm{km\,s^{-1}}$, respectively, at the end of the fitting. The time 16:00~UT is marked with vertical dotted lines to compare the eruptions of the two prominences.
    %The turns from the slow-rise phase to the fast-rise phase are generally determined when the velocity of the exponential part equals to that of the linear part \citep{2013ApJ...769L..25C}. However, the calculated turning times for both PRO-S and PRO-N are before 14:00~UT, much earlier than their eruptions. It could be caused by a small rising velocity in the slow-rise phase, or there is no turn during their accelerations. Hence we just use 16:00~UT as a reference time to compare the eruptions of the two prominences (vertical dotted lines).  \added{The turns from the slow-rise phase to the fast-rise phase are generally determined when the speed of the exponential part equals to that of the linear part \citep{2013ApJ...769L..25C}. However, in our case, the calculated turning times are at least 3 hours earlier than the apparent eruptions due to a small speed of the linear part.}
    % which are calculated to be 13:51~UT and 11:17~UT for PRO-S and PRO-N, respectively

    From Figure~\ref{fig-slice}b--c, it is seen that PRO-S starts erupting before 16:00~UT and PRO-N rises up following PRO-S. Though no CME is found
    following their eruptions, PRO-S is fully destroyed but the bottom part of
    PRO-N survives. In Figure~\ref{fig-slice}d, the lower (south) structure is
    PRO-S and the upper (north) one is PRO-N. During the PRO-S eruption, part of
    PRO-N mass flows northwards in the plane-of-sky (PoS); after PRO-S eruption,
    PRO-N mass flows oppositely. The motion of PRO-N is likely to be associated with
    the PRO-S eruption.
    % We will see that the two eruptions are very different in next Section.

    \subsection{Phenomena Related to PRO-S Eruption}
    \begin{figure}
        \centerline{\includegraphics[width=1\textwidth,clip=]{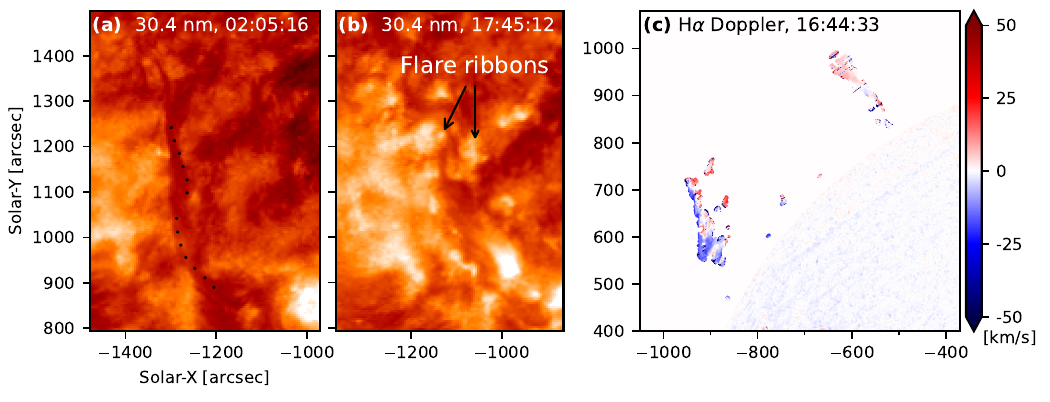}}
        \caption{\label{fig-eui_proS}Clues of causes of PRO-S eruption. (a)--(b) EUI $30.4\,\mathrm{nm}$ snapshots on 26 October 2022 at 2:05 and 17:45~UT, respectively. (c) H$\alpha$ Doppler map on 26 October 2022 at 16:44~UT.}
    \end{figure}
    \begin{figure}
        \centerline{\includegraphics[width=1\textwidth,clip=]{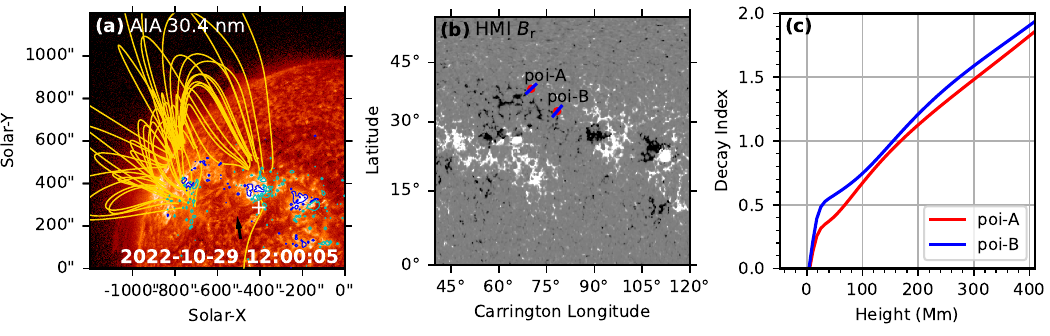}}
        \caption{\label{fig-pfss}Potential field configuration and decay index. (a) AIA $30.4\,\mathrm{nm}$ snapshots on 29 October 2022 at 12:00~UT and extrapolated magnetic field configuration, where only closed field is plotted as golden lines. Blue and cyan contours represent $B_\mathrm{los}$ of -100 and $100\,\mathrm{G}$, respectively. (b) $B_\mathrm{r}$ Carrington map from HMI observations.  (c) Decay index along height.}
    \end{figure}
    \begin{figure}
        \centerline{\includegraphics[width=1\textwidth,clip=]{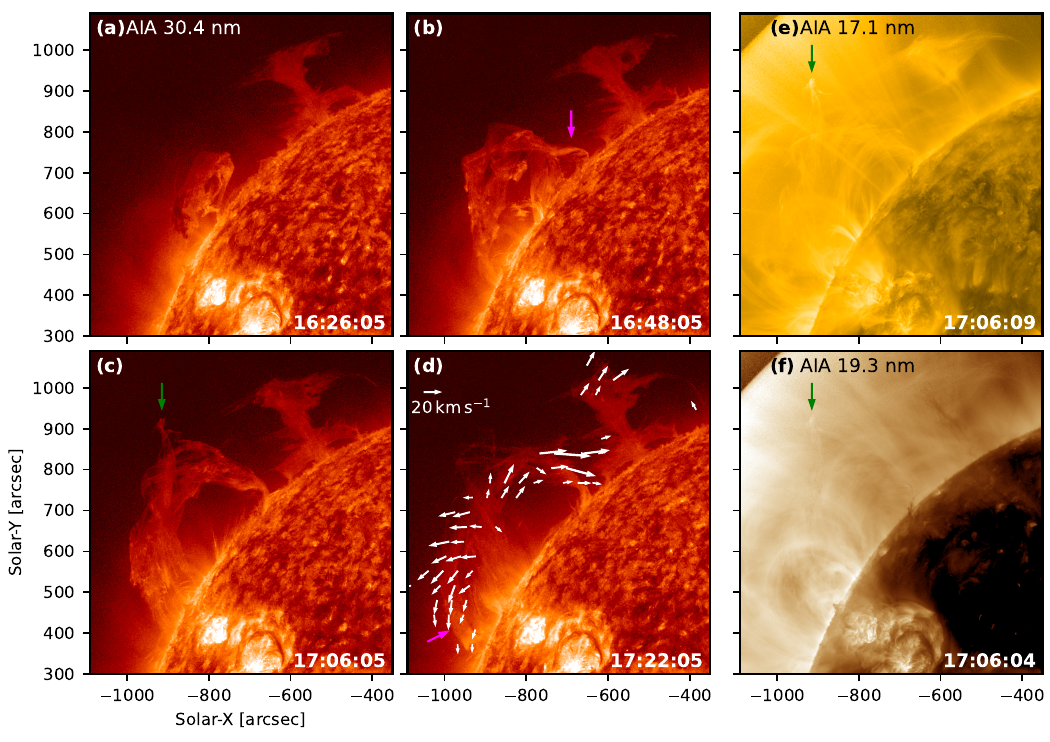}}
        \caption{\label{fig-aia_proS}Process of the PRO-S failed eruption. (a)--(d) AIA $30.4\,\mathrm{nm}$ images. White arrows in (d) represent flow in the plane-of-sky. (e) AIA $17.1\,\mathrm{nm}$ image. (f) AIA $19.3\,\mathrm{nm}$ image. Observation times are marked at lower-right corner in each panel.}
    \end{figure}
    
    As mentioned in Section~\ref{s:intro}, the initiation and failing of an
    eruption can be related to a kink instability, a torus instability, and
    reconnection below or/and around the flux rope, in addition to external triggers.
     We explore the eruption of PRO-S in relation with these mechanisms.

    Figure~\ref{fig-eui_proS}a--b shows snapshots of EUI $30.4\,\mathrm{nm}$
    maps of PRO-S. Around 14 hours before the PRO-S eruption
    (Figure~\ref{fig-eui_proS}a), a flux rope with a twist number of
    nearly 2 is seen (delineated with dotted curves), and the helicity is negative.
    Its southern leg roots in the periphery of an active region (AR), and the
    northern leg finishes in bifurcated ends. During the PRO-S eruption
    (Figure~\ref{fig-eui_proS}b), flare ribbons are seen underneath.
    Figure~\ref{fig-eui_proS}c shows a Doppler map from CHASE H$\alpha$
    observations during the PRO-S eruption. A single Gaussian fitting is used to derive the Doppler speeds for simplification \citep{2021RAA....21..222X}, hence the derived speeds should be treated as an averaged result, or represent the motions of the main component if there are multi-peaks in one H$\alpha$ profile. Relative to the PRO-S leg axis, the
    left part is mainly blue-shifted and the right part is red-shifted with
    velocities within around $\pm 17\,\mathrm{km\,s^{-1}}$. Compared with the
    negative helicity shown in Figure~\ref{fig-eui_proS}a and assuming that the
    footpoint is tied, the Doppler map in Figure~\ref{fig-eui_proS}c indicates
    an untwisting motion.

    The threshold of the torus instability is usually given by a critical decay index.
    The decay index $n$ is defined by
    \begin{equation}
        n = -\frac{\mathrm{dln}B_\mathrm{ep}(R)}{\mathrm{dln}R}\,,
    \end{equation}
    where $R$ denotes the distance to the center of an approximately toroidal flux rope, and $B_\mathrm{ep}$ denotes the external poloidal field \citep{2023ApJ...951L..35C}. For simplification, we use height to replace $R$ and $B_\mathrm{ep}$ is assumed to be nearly perpendicular to the polarity inversion line (PIL) based on a potential field approach. Through comparing $B_\mathrm{los}$ of PHI and HMI, the magnetic field around PRO-S does not vary significantly in the following days. Hence the synoptic Carrington magnetogram of the normal component ($B_\mathrm{r}$) from HMI is used to compute the potential field with the \emph{pfsspy} package \citep{2020JOSS....5.2732S}. The extrapolated potential field configuration is shown in Figure~\ref{fig-pfss}a over an AIA $30.4\,\mathrm{nm}$ map, where only closed field lines are plotted. The blue and cyan contours mark HMI $B_\mathrm{los}$ of -100 and $+100\,\mathrm{G}$, respectively. Though PRO-S erupts on 26 October, the filament channel is clearly seen (pointed by a black arrow, no corresponding filament is seen in the H$\alpha$ image), which is enveloped by a series of loops. The south footpoint of PRO-S lies on a positive polarity, it can be seen from the overlaid $B_\mathrm{los}$ contours, and could also be derived from the fact that a negative helicity structure locates along a dextral channel \citep{1994ASIC..433..303M}. The decay index is calculated at two points around the middle point of the projected PRO-S axis along the PIL, marked as poi-A and poi-B in Figure~\ref{fig-pfss}b, and its changes with height are shown in Figure~\ref{fig-pfss}c. From Figure~\ref{fig-slice}b, the axis of PRO-S reaches a height of around $185\,\mathrm{Mm}$ before its disappearance, which should be the lower limit considering the projection effect. The decay index at $185\,\mathrm{Mm}$ is $\approx 1$, reaching the critical value of the torus instability for a straight current channel, but smaller than 1.5 for a circular shape \citep{2010ApJ...718.1388D}. Considering the errors of measurements in this work and the limitations in models, we cannot say whether PRO-S keeps torus stable during the rising process.

    Signatures of magnetic reconnection between PRO-S and the ambient coronal loops
    are explored in Figure~\ref{fig-aia_proS}, which mainly shows sequential images
    of PRO-S failed eruption in AIA $30.4\,\mathrm{nm}$. The first signature is the
    appearance of the new legs formed during the eruption. The two
    new legs are marked with magenta arrows in Figure~\ref{fig-aia_proS}b and
    d. They are filled with PRO-S dropping mass. The second signature is that the rising PRO-S is not regular and a
    ``peak'' is marked (green arrow) in Figure~\ref{fig-aia_proS}c, which is also
    bright in AIA 17.1 and $19.3\,\mathrm{nm}$ images, suggesting that the plasma
    is heated to coronal temperature. However, because the emission of the PRO-S
    eruption is relatively weak and no X-ray flare is detected by the GOES, there
    is no significant brightening around PRO-S. A brightening in X-rays and EUV is
    usually used as an indication of magnetic reconnection
    \citep{2003ApJ...595L.135J,2012A&A...548A..89N,2016ApJ...832..106H,2023ApJ...959...67C}.
    % Writhing motion is necessary for magnetic reconnection with overlying field, if there is no quadrupolar field.
    % PRO-S has an orientation of southeast to northwest from EUI view (Figure~\ref{fig-eui_phi}), consistent with the AIA view in Figure~\ref{fig-aia_proS}a. Comparing Figure~\ref{fig-aia_proS}b with a, the erupting PRO-S rotates clockwise in top view.
    % A clockwise rotation is consistent with the transforming of twist into writhe for PRO-S.
    %To keep the conversation of magnetic helicity, a clockwise writhing is needed during the untwisting motion of PRO-S. 
    In Figure~\ref{fig-aia_proS}d, the optical flows (marked with white arrows,
    calculated using two $30.4\,\mathrm{nm}$ images with an interval of 2~minutes)
    suggest that the PRO-S mass is mainly dropping along the legs and the rising
    almost stops.

    \subsection{Phenomena related to PRO-N eruption}
    \begin{figure}
        \centerline{\includegraphics[width=1\textwidth,clip=]{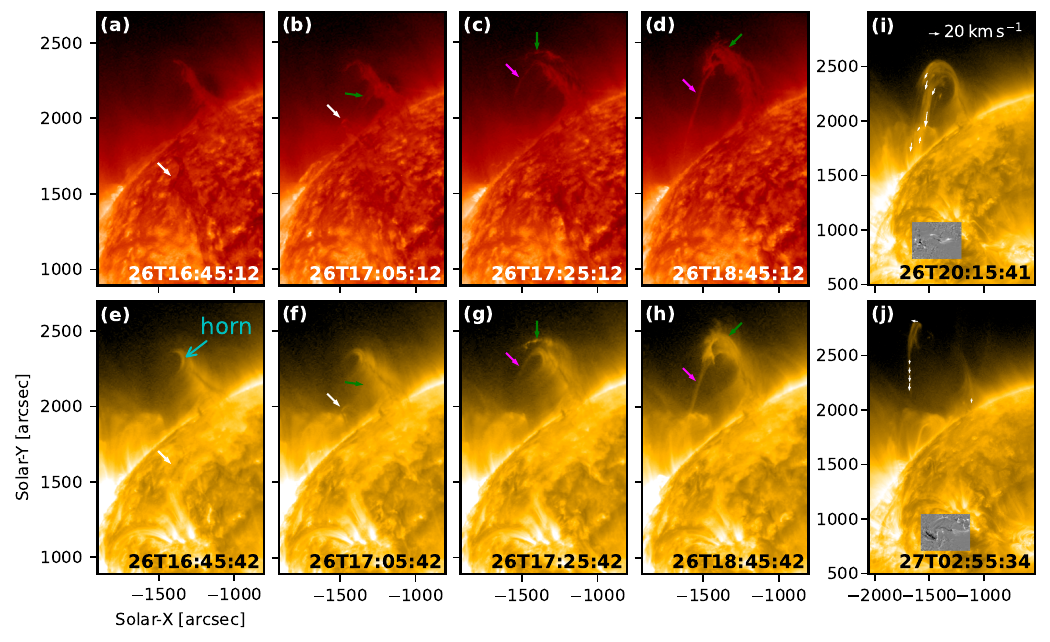}}
        \caption{\label{fig-eui_proN}Time sequence images of EUI $30.4\,\mathrm{nm}$ (a-d) and $17.4\,\mathrm{nm}$ (e-j) showing PRO-N mass drainage. In (i)--(j), the white arrows represent flows, and the grey insets are difference images.}
        %White arrows point to the erupting PRO-S, purple arrows to the fast-rising mass around PRO-N, and the magenta arrows to PRO-N mass flow. 
    \end{figure}
    The end of PRO-N is in the form of
    mass drainage after PRO-S failed eruption. Figure~\ref{fig-eui_proN} shows
    the draining process in EUI 30.4 and $17.4\,\mathrm{nm}$. Among
    the panels, the white arrows in panels a, b, e, and f point to the erupting
    PRO-S; the green arrows point to a cluster of mass that rises faster than PRO-N main body, and stops around the horn-like structure of PRO-N; and the magenta arrows point the mass draining along the horn. Horn-like structures refer to curved extensions that protrude from the top of quiescent prominences into the cavities \citep{2015ApJ...807..144S}.
     In our case, the horn is clearly seen in $17.4\,\mathrm{nm}$ images, and the curvature gets smaller during the
    PRO-S rising. After PRO-S failed eruption, the flow along the horn gets
    significant.
    Figure~\ref{fig-eui_proN}i--j shows violent drainage of PRO-N mass
    along the horn (white arrows mark the optical flow) and the brightening on the disk
    (difference image of the inset). Due to
    the projection effect, we cannot identify the other end of the flux tube that
    connects the PRO-N horn. However, the brightening on the disk has a good
    temporal correlation with the violent mass drainage (see movie\_c.mp4).
    Hence, we suspect that the PRO-N horn extends southwards and overlies the
    filament channel where PRO-S is located in before the eruption.

    %%%%%%%%%%%%%%%%%% Discussion %%%%%%%%%%%%%%%%%%%%%%%
    \section{Discussion and Conclusion}\label{s:dis}
    Through the time-distance diagrams, we have seen that the motion of PRO-N is closely related to PRO-S eruption: PRO-N rises up along with PRO-S, and PRO-N mass moves northwards during PRO-S expansion and moves back after PRO-S failed eruption.

    Then we study the causes of PRO-S failed eruption, and mainly explored the evidence of whether kink instability, torus instability, flare reconnection, and reconnection with ambient magnetic field play roles. PRO-S has a twist number of nearly 2 and shows an untwisting motion during the eruption (Figure~\ref{fig-eui_proS}). Therefore, we suggest that a kink instability happens. A trigger is generally necessary to excite the kink instability, and the trigger resulting in loss of equilibrium could be an oscillation and mass drainage \citep{2014ApJ...790..100B,2020ApJ...898...34F}, magnetic reconnection in the rope system \citep{2023FrASS...984678Z} including the one happening in the breakout model \citep{2023ApJ...943..156K}. The decay index is calculated using an HMI Carrington magnetogram and is found to be around 1 at the height of PRO-S, just reaching the lower limit for the torus instability (1 to 1.5). Hence it is possible that the external field, no matter poloidal or toroidal field once PRO-S rotates, plays a role in confining the eruption. There are some shortcomings if a potential field model is used as an approximation of the real external field. In addition to the problems mentioned in \citet{2023arXiv231207406Z}, the ambient magnetic field may be disturbed by frequent activity. Before the event we study, a faint but large filament eruption occurs around 1:00~UT on 26 October 2022 (see movie\_c.mp4). Flare reconnection under a flux rope is thought to be the main cause of explosive CME acceleration \citep{2012ApJ...760...81K}. We observe flare ribbons but no post-eruption loops, which suggests that the flare reconnection happens but is very weak during PRO-S eruption. The reconnection between PRO-S and the ambient field is also explored, and new legs filled with dropping prominence mass probably support the occurrence of magnetic reconnection. However, due to the low emission of the eruption (no GOES X-ray flare), the reconnection location cannot be identified from X-ray or EUV radiation. A difference of PRO-S from the generally reported confined eruptions \citep{2023ApJ...959...67C} is that no clear deceleration is detected before the disappearance of PRO-S along the rising trajectory (Figure~\ref{fig-slice}b). We suggest that PRO-S is destroyed by the magnetic reconnection between PRO-S and the overlying field before an obvious deceleration. Considering the observed untwisting motion and expected writhing motion (to keep the conservation of magnetic helicity), it is speculated that the magnetic reconnection happens (see the fourth regime for failed eruptions in Section~\ref{s:intro}).    
    %  The vanishment of PRO-S could be caused by reconnection with ambient field, or PRO-S emission declines fast enough before the PRO-S deceleration.

    A horn-like structure is at the top of PRO-N, along which PRO-N mass drains southwards. Brightening occurs on some regions on the solar disk, and these regions may connect with PRO-N horn through flux tubes. Prominence horn structures are generally thought to be flux ropes or hyperbolic flux tubes holding prominences \citep{2012ApJ...758...60F,2015ApJ...807..144S}. From the EUI $30.4\,\mathrm{nm}$ image in Figure~\ref{fig-eui_phi}a, RPO-S has an orientation southwest to northeast. We suspect the flux tube connecting PRO-N horn is over PRO-S, which could explain both the rising up of PRO-N along with PRO-S, and mass drainage after PRO-S failed eruption due to evacuation under the flux tube. Once the mass flow starts, it leads to a catastrophic mass drainage of PRO-N because of a siphon effect.
    Plasma gravity plays an important role in stabilizing a prominence and continuous mass drainage may initiate an eruption \citep{2020ApJ...898...34F,2023ApJ...956..106W}. From Figure~\ref{fig-eui_proN}i--j, PRO-N rises during the violent mass drainage but does not erupt.
    
    The relation between PRO-S and PRO-N is different from the flux rope systems of double-decker above the same PIL \citep{2012ApJ...756...59L} or side-by-side over different PILs. 
    Due to the limitation of observations, we cannot identify exactly the location of the flux tube that connects with PRO-N horn.
    Horn structures, not rare for quiescent prominences \citep{2016ApJ...827L..33W}, are a part of long flux tubes, including both flux ropes and sheared arcades. Prominence and flux tubes that connect with the horns compose the large scale prominence-cavity systems. They suffer frequent disturbance because of activity underneath including jet and filament eruptions \citep{2021ApJ...923L..10C}. From EUI observations, PRO-S is likely to locate below the flux tube that holds PRO-N.
    % PRO-S is likely to locate below or beside the prominence-cavity system of PRO-N, and have a close relation with the flux tube that hangs PRO-N.
    %Horn structures are not rare for quiescent prominences, they suffers frequent disturbance from the activities underneath including jets and  

    In summary, the sympathetic eruptions in this work could be divided into three stages:
    \begin{enumerate}
        \item PRO-S rises due to a helical kink instability (PRO-S is twisted in EUI $30.4\,\mathrm{nm}$ image and an untwisting motion is observed in the H$\alpha$ Doppler map). Magnetic reconnection occurs under PRO-S (flare ribbons) and provides a positive feedback to the eruption. Because the flux tube that holds PRO-N is over PRO-S, PRO-N also rises up and is pushed away from PRO-S due to the expansion of the latter (time-distance diagrams in Figure~\ref{fig-slice}).
        \item PRO-S reconnects with the overlying magnetic field (new legs appear in Figure~\ref{fig-aia_proS}), and the upward hoop force decreases along with the decay of the poloidal field of PRO-S. It is possible that the external magnetic field plays a role in confining the eruption. 
        \item PRO-N mass flow along the horn starts due to the evacuation after PRO-S failed eruption, which leads to the catastrophic mass drainage due to siphon effect (Figure~\ref{fig-eui_proN}). In the end only the bottom part remains.
    \end{enumerate}

    % Due to the limitation of observations, some assumptions are made to explain the failed eruption of PRO-S.
    Mass draining is a usual form in which prominences end their life, and this work suggests that catastrophic mass draining may result from a filament eruption underneath.

    %%%%%%%%%%%%%%%%%%%%%%%%%%%%%%%%%%%%%%%%%%%%%%%%%%%%%%%%%%%%%%%%%%%%%%%%%%%
    \section*{Acknowledgements}
        The authors thank the reviewer very much for the valuable comments and suggestions that helped improve the manuscript. The ASO-S is supported by the Strategic Priority Research Program on Space Science, Chinese Academy of Sciences. The CHASE mission is supported by China National Space Administration. SDO is a mission for NASA's Living with a Star Program. Solar Orbiter is a space mission of international collaboration between ESA and NASA, operated by ESA. The EUI instrument was built by CSL, IAS, MPS, MSSL/UCL, PMOD/WRC, ROB, LCF/IO with funding from the Belgian Federal Science Policy Office, the Centre National d’Etudes Spatiales (CNES), the UK Space Agency (UKSA), the Bundesministerium f$\mathrm{\ddot{u}}$r Wirtschaft und Energie (BMWi), and the Swiss Space Office (SSO). This work is supported by the National Natural Science Foundation of China (NSFC) 12233012, the Strategic Priority Research Program of the Chinese Academy of Sciences XDB0560000, National Key R\&D Program of China 2022YFF0503003 (2022YFF0503000), the mobility program (M-0068) of the Sino-German Science Center, NSFC (grant Nos. 11973012, 11921003, 12203102, 12103090, 12333009), the Jiangsu Funding Program for Excellent Postdoctoral Talent, the Youth Talent Cultivation Program of the PMO E2ZC161111. 

\bibliographystyle{spr-mp-sola}
\bibliography{PE}
\end{document}